\documentclass[aps, prl,twocolumn,superscriptaddress]{revtex4-1}%groupedaddress
\usepackage{graphicx}
\usepackage{hyperref}
\usepackage{mathtools}
\DeclarePairedDelimiter\bra{\langle}{\rvert}
\DeclarePairedDelimiter\ket{\lvert}{\rangle}
\DeclarePairedDelimiterX\braket[2]{\langle}{\rangle}{#1 \delimsize\vert #2}

\begin{document}
%\preprint{}

\title{Readout Of Singlet-Triplet Qubits At Large Magnetic Field Gradients}


\author{Lucas A. Orona}
\affiliation{Department of Physics, Harvard University, Cambridge, MA 02138, USA}
\author{John M. Nichol}
\affiliation{Department of Physics, Harvard University, Cambridge, MA 02138, USA}
\affiliation{Department of Physics and Astronomy, University of Rochester, Rochester, NY 14627, USA}
\author{Shannon P. Harvey}
\author{Charlotte G. L. B{\o}ttcher}
\affiliation{Department of Physics, Harvard University, Cambridge, MA 02138, USA}
\author{Saeed Fallahi}
\author{Geoffrey C. Gardner}
\affiliation{Department of Physics and Astronomy, Purdue University, West Lafayette, IN 47907, USA}
\author{Michael J. Manfra}
\affiliation{Department of Physics and Astronomy, Purdue University, West Lafayette, IN 47907, USA}
\affiliation{School of Materials Engineering, Purdue University, West Lafayette, IN 47907, USA}
\affiliation{Birck Nanotechnology Center, Purdue University, West Lafayette, IN 47907, USA}
\affiliation{School of Electrical and Computer Engineering, Purdue University, West Lafayette, IN 47907, USA}
\author{Amir Yacoby}
\affiliation{Department of Physics, Harvard University, Cambridge, MA 02138, USA}

%\thanks{}
%\affiliation{1-Department of Physics, Harvard University, Cambridge, MA 02138, USA 2- Department of Physics, Rochester University, Rochester, NY 14627, USA 3-Department of Physics and Astronomy, Purdue University, West Lafayette, IN 47907, USA 4-Birck Nanotechnology Center, Purdue University, West Lafayette, IN 47907, USA 5-School of Materials Engineering, Purdue University, West Lafayette, IN 47907, USA 6-School of Electrical and Computer Engineering, Purdue University, West Lafayette, IN 47907, USA}


\date{\today}

\begin{abstract}
Visibility of singlet-triplet qubit readout is reduced to almost zero in large magnetic field gradients due to relaxation processes.  Here we present a new readout technique that is robust against relaxation and allows for measurement when previously studied methods fail. This technique maps the qubit onto spin states that are immune to relaxation using a spin dependent electron tunneling process between the qubit and the lead.  We probe this readout's performance as a function of magnetic field gradient and applied magnetic field, and optimize the pulse applied to the qubit through experiment and simulation. 

\end{abstract}

\pacs{}
%\keywords{}

\maketitle

%\section{Introduction}
%\subsection{}
%\subsubsection{}

Electron spins in semiconductors\cite{Loss,Koppens,Pioro,Kim,Eng} are one promising path to quantum computing because of their scalability and long coherence times\cite{Veldhorst,Muhonen,Saeedi}.  Single qubit gate fidelities exceed 99.8$\%$ in single electron spin qubits\cite{Muhonen2} and 99$\%$ in singlet-triplet(S-T) qubits\cite{Nichol}.  S-T qubits \cite{Petta,Foletti2,Shulman} have recently demonstrated two qubit gate fidelities of 90$\%$ by using large magnetic field gradients\cite{Nichol}, $\Delta B_{z}$, to diminish the effects of charge noise\cite{Dial} and increase coherence times.  However, in the presence of $\Delta B_{z} >$ 400 MHz relaxation through coupling to other states reduces readout visibility to almost zero\cite{Barthel}. 

Here we report a new readout scheme that provides readout contrast at large gradients and demonstrate that it has superior performance to previously published methods \cite{Petta,Studenikin} for $\Delta B_z>$ 500 MHz.  This method is robust up to at least $\Delta B_z$ = 900 MHz, the largest magnetic field we could generate, and should continue to function in much larger $\Delta B_{z}$.  S-T qubits have previously been read out by mapping the qubit states on different charge configurations\cite{Petta}.  However, large gradients enable transitions between the qubit states during measurement, leaving both in the same charge configuration and diminishing contrast.  Our technique adds a step before measurement that shelves the qubit states into alternate spin states that do not have relaxation pathways enabled by $\Delta B_{z}$, restoring the ability to map each spin state onto a distinct charge configuration.  This method relies on a  spin-dependent tunneling between the qubit and the surrounding two dimensional electron gas (2DEG).

To optimize this process, we have measured the visibility of our readout as a function of $\Delta B_{z}$,  the voltage applied during shelving and its duration, and magnetic field, $B$.  We have also developed a simple model for this readout and used it to simulate our experiments, finding strong agreement with the data.  The model we introduce is applicable to other varieties of spin qubits, including single spin\cite{Elzerman, Morello}, hybrid qubit\cite{Kim2}, and donor based S-T qubit\cite{Broome} and latched readout methods\cite{Harvey, Nakajima} that also rely on tunneling between the qubit and a Fermi sea.  This readout technique is general to any host material, and source of $\Delta B_{z}$ and to schemes that use S-T readout for single spin qubits\cite{Fogarty}.

%\section{Device}

\begin{figure}
	\includegraphics[scale=.42]{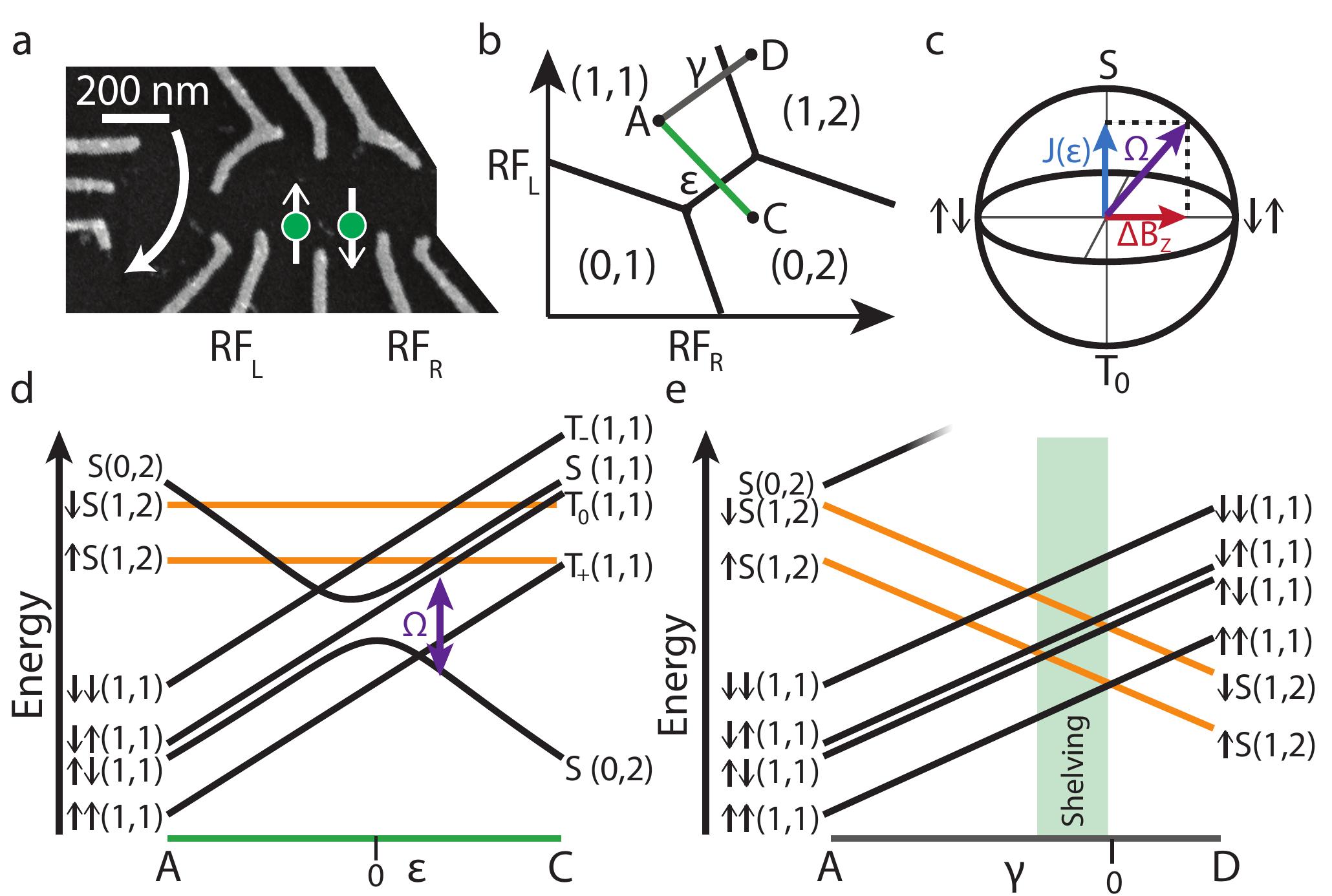}%
	\caption{(a) SEM image of the device.  Electron positions are approximated with  green circles.  The sensor quantum dot is shown with a white arrow.  (b) Charge stability diagram of the qubit.  In the experiment voltages are either applied equally, $\gamma$, or oppositely, $\epsilon$, to the RF gates.  (c) Bloch sphere of the qubit showing the eigenstates of J, $\Delta B_{z}$ and total splitting $\Omega$. (d) The energies of relevant states along the $\epsilon$ curve in b.  (e) The energies of relevant states along the $\gamma$ curve in b.  In both (d) and (e) black and orange curves represent the energies of two and three electron states respectively.}
\end{figure}
We study S-T qubits formed from two electrons trapped in an electrostatic gate defined double quantum dot in the 2DEG of GaAs shown in Figure 1a.  We use the pair of numbers (L,R) to represent the number of electrons in the left and right dots respectively.  The logical subspace for the qubit is made up of the singlet, \begin{math}\ket{S}=(\ket{\uparrow\downarrow}-\ket{\downarrow\uparrow})/\sqrt{2}\end{math}, and triplet, \begin{math}\ket{T_{0}}=(\ket{\uparrow\downarrow}+\ket{\downarrow\uparrow})/\sqrt{2}\end{math}, states where the arrows represent the electron spin in the left and right dot respectively.  The Hamiltonian for this system is given by \begin{math} H=\Delta B_{z}\sigma_{x}+J(\epsilon)\sigma_{z} \end{math}\cite{Petta}.  The exchange interaction, $J(\epsilon)$, splits S from T$_{0}$ and is controlled by the detuning, $\epsilon$, and the energy splitting between $\uparrow\downarrow$ and $\downarrow\uparrow$ is controlled by $\Delta B_{z}$.  We call the magnitude of the Hamiltonian \begin{math}\Omega(\epsilon)=\sqrt{\Delta B_{z}^{2}+J(\epsilon)^{2}} \end{math}, as shown in Figure 1c,d.  We note that the nature of the qubit’s ground (excited) state changes from being S (T$_{0}$) in (0,2) to $\uparrow\downarrow$ ($\downarrow\uparrow$) in (1,1). 

For all experiments in this work, $\Delta B_{z}$ is produced by the hyperfine interaction with the nuclei, which is controlled through dynamic nuclear polarization (DNP)\cite{Foletti} applied prior to every experimental run.  The qubit is manipulated by applying voltage pulses to the gates labeled RF$_{L}$ and RF$_{R}$ in Figure 1a.  The total number of electrons in the double dot is controlled by
 \begin{math}
 \gamma=($RF$_{L}+$RF$_{R})/2
 \end{math}
 and the distribution of these between the right and left dot is controlled by
 \begin{math}
 \epsilon=$RF$_{L}-$RF$_{R}
 \end{math}
 , shown in Figure 1b.  We define \begin{math}
 \gamma=0
 \end{math}
 to be the transition from the (1,1) to the (1,2) region, as shown in Figure 1e.  The qubit's charge state is measured using an additional neighboring quantum dot\cite{Barthel2}.  
 
%\section{Shelving Mechanism}

\begin{figure}
	\includegraphics[scale=.45]{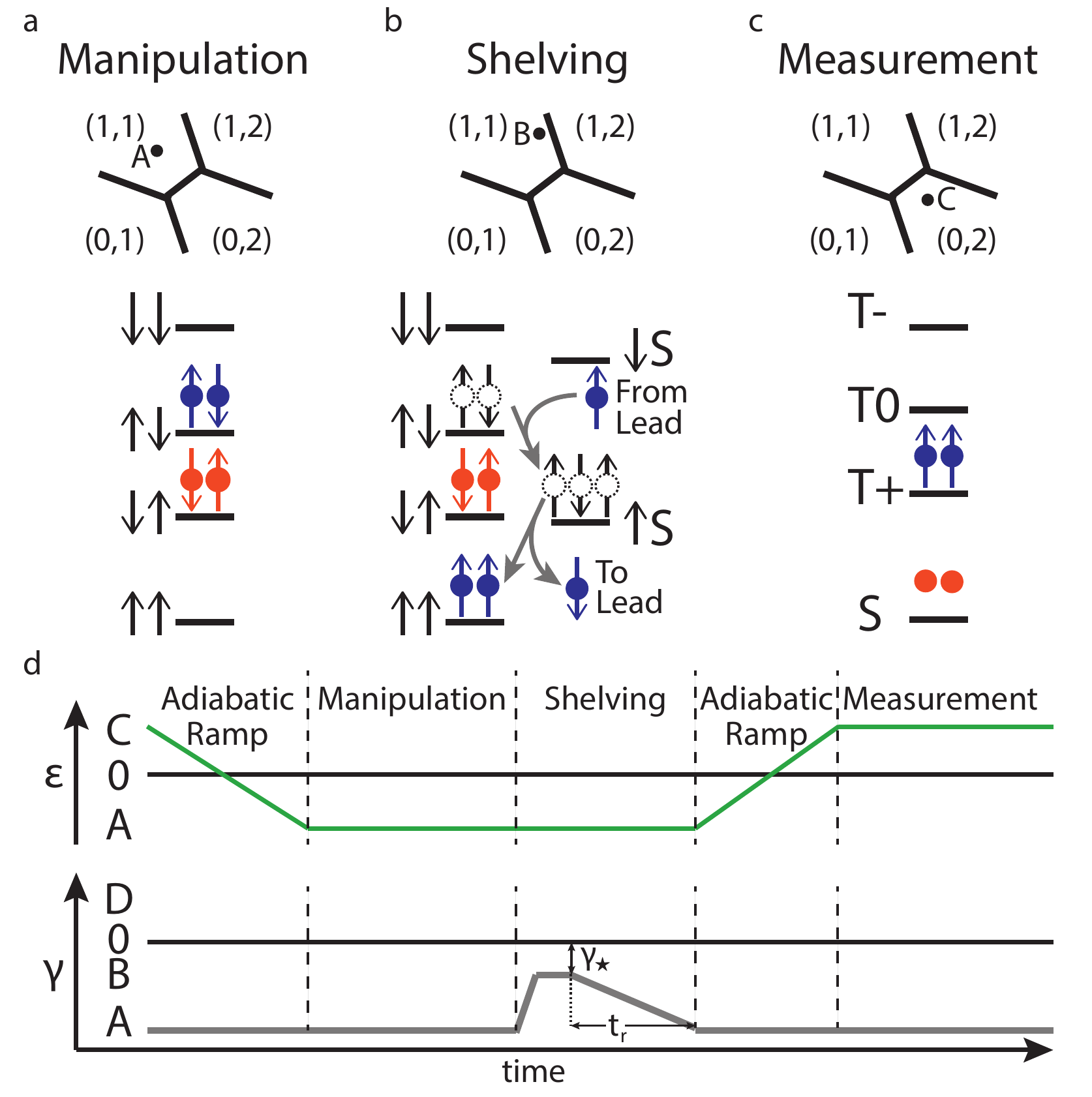}%
	\caption{(a-c) Position in the charge stability diagram and occupation of quantum dot states after (a)manipulation, (b) shelving, and (c) measurement.  For (a-c) the qubit's excited state is shown in blue while the qubit ground state is shown in red. (a)  After manipulation, the qubit is in its logical subspace, $\uparrow\downarrow$ and $\downarrow\uparrow$.  (b) Grey arrows represent the transitions required for shelving to occur.  Filled circles show states that are occupied at the end of process while dotted circles show states that are empty.  (c)  State occupation at the measurement position.  The T$_+$ and S states cannot be mixed by $\Delta B_z$.  (d) Pulse sequence.  Values of $\epsilon$ and $\gamma$ during different steps of qubit operation.  The shelving position, set by $\gamma_{\star}$ and ramp time, t$_r$, to (1,1) are optimized in Figure 4.}
\end{figure}
We manipulate our qubits deep at Position A, shown in Figure 2a, where the two spins are well isolated so that the ground state is $\downarrow\uparrow$(1,1) and the excited state is $\uparrow\downarrow$(1,1).  In previous work S-T qubits were read out through spin blockade by adiabatically ramping the qubit from deep in (1,1) to the measurement point in the (0,2) region.  This point is chosen so that S is in (0,2) but T$_{0}$ is spin blockaded to remain in (1,1) because excited energy levels of the quantum dot are energetically inaccessible.  This readout process maps $\downarrow\uparrow$(1,1) to S(0,2) and $\uparrow\downarrow$(1,1) to T$_{0}$(1,1) so that the distinct charge configurations can be used to measure the qubit's spin state.  However, this style of readout is vulnerable because at the measurement point $\Delta B_z$ mixes T$_{0}$(1,1) with the excited S(1,1) state, which decays to S(0,2) on time scales much shorter than the measurement time\cite{Barthel}.  When this transition occurs, there is no readout contrast because both qubit states have the same charge configuration.  The rate of transition from T$_{0}$(1,1) to the excited S(1,1) state increases with $\Delta B_{z}$, meaning that this method has a measurement fidelity that decreases with increasing $\Delta B_{z}$.

To overcome readout failure at large $\Delta B_{z}$ we developed a new readout technique that shelves the qubit states into readout states which do not have relaxation pathways enabled by $\Delta B_z$.  This new method maps $\downarrow\uparrow$(1,1) to S(0,2) and $\uparrow\downarrow$(1,1) to T$_{+}$(1,1).  For the remainder of the work, we will refer to this as the the T$_{+}$ readout method.  We achieve the desired mapping by using tunneling between the right quantum dot and the 2DEG to change the qubit's spin state.  The qubit is tuned so that the left dot is isolated from the lead and the other dot.  The shelving process is shown in Figure 2a-c and begins deep in (1,1), at Point A.  After manipulation, the qubit is brought to Point B, where $\gamma=\gamma_{\star}$, which is chosen so the required transitions are energetically favorable, as shown in Figure 2b.   At this point, electrons can only tunnel in and out of the right dot, enabling the transition from $\uparrow\downarrow$(1,1) to $\uparrow$S(1,2) by a spin $\uparrow$ electron tunneling in.  The transition from $\downarrow\uparrow$(1,1) to $\uparrow$S(1,2) is blocked because there is no mechanism to change the spin in the left dot.  $\uparrow$S(1,2) decays to $\uparrow\uparrow$(1,1) by a spin $\downarrow$ electron tunneling from the right dot to the lead.  After allowing the qubit to fully transition, the voltages are adiabatically changed back to Point A over a time t$_{r}$ and then brought to Point C, the same measurement point as in the spin blockade method.  The charge state is then measured with S(0,2) corresponding to the ground state, $\downarrow\uparrow$(1,1), and T$_{+}$(1,1) corresponding to the excited state, $\uparrow\downarrow$(1,1).

This technique also enables us to measure the direction of $\Delta B_{z}$.  We have described this mechanism assuming a specific directionality for $\Delta B_{z}$ but it functions with the opposite orientation as well.  Flipping the direction of $\Delta B_{z}$ causes $\uparrow\downarrow$(1,1) to be the ground state and $\downarrow\uparrow$(1,1) to be the excited state.  This readout still maps $\uparrow\downarrow$(1,1) to T$_{+}$(1,1) while $\downarrow\uparrow$(1,1) is initially mapped to T$_{0}$(1,1) and quickly decays to the S(0,2) charge state through the mechanism previously described.  This inverts the charge signal we measure from the qubit ground state, allowing for a direct measurement of the direction of $\Delta B_{z}$.  In these experiments, $\Delta B_{z}$ is oriented as in the second regime because DNP is more effective when pumping with T$_+$ than S, as detailed in the Supplementary Materials.

These readout techniques are sufficient for full qubit state tomography because we are able to pair them with high fidelity single qubit gates.  We can measure along any axis by performing the proper rotations so that the states along the desired axis are mapped onto $\uparrow\downarrow$(1,1) and $\downarrow\uparrow$(1,1).

%\section{Theoretical Model}

We have constructed a simple model for the T$_{+}$ method that captures the experimental trends that we observe and offers intuition for this technique's behavior.  To determine the equilibrium populations of all the different quantum dot states, we have calculated the transition rates between all pairs of states using Fermi's golden rule to compute the tunneling rates of electrons between the qubit and the 2DEG.  We find the following transition rates, $\Gamma_{ij}$ between the (1,1) states, $i$, and the (1,2) states, $j$, and the reverse, $\Gamma_{ji}$:

\begin{align}
\Gamma_{ij} &=\frac{2\pi}{\hbar}|\bra{j}\tau\ket{i}|^{2}f(\Delta E_{ij},T,\mu)\rho_{f}\\
\Gamma_{ji} &=\frac{2\pi}{\hbar}|\bra{i}\tau\ket{j}|^{2}(1-f(-\Delta E_{ji},T,\mu))\rho_{f}
\end{align}

Here $\hbar$ is the reduced Planck constant, $\tau$ is the tunneling term between the right quantum dot and 2DEG, f is the Fermi-Dirac distribution, $\Delta E_{ij}=E_{j}-E_{i}$ is energy difference between $i$ and $j$, $T$ is the electron temperature, and $\mu$ and $\rho_{f}$ are the chemical potential and density of states of the 2DEG.  $\Delta$E$_{ij}$ is controlled by $\epsilon$, $\gamma$, $\Delta B_{z}$, and $B$.  Transitioning between states with different numbers of electrons requires an electron tunneling to or from the lead with an energy that compensates for any change to the qubit's energy.  The Fermi-Dirac distribution  dictates the number of electrons and holes available for $\Gamma_{ij}$ and $\Gamma_{ji}$ respectively, which governs the rates.  This means that the transition rates from states with lower energy to higher energy are suppressed because they require an excited electron or hole to donate the energy difference.   We note that many rates are 0 due to spin conservation, suppressing transitions between states with incompatible spin configurations. We use these rates to simulate the transitions that occur during T$_+$ readout so that we can perform simulations while varying the same parameters as we do experimentally.  Details are included in the Supplementary Materials.

%\section{Results}

\begin{figure}
	\includegraphics[scale=.53]{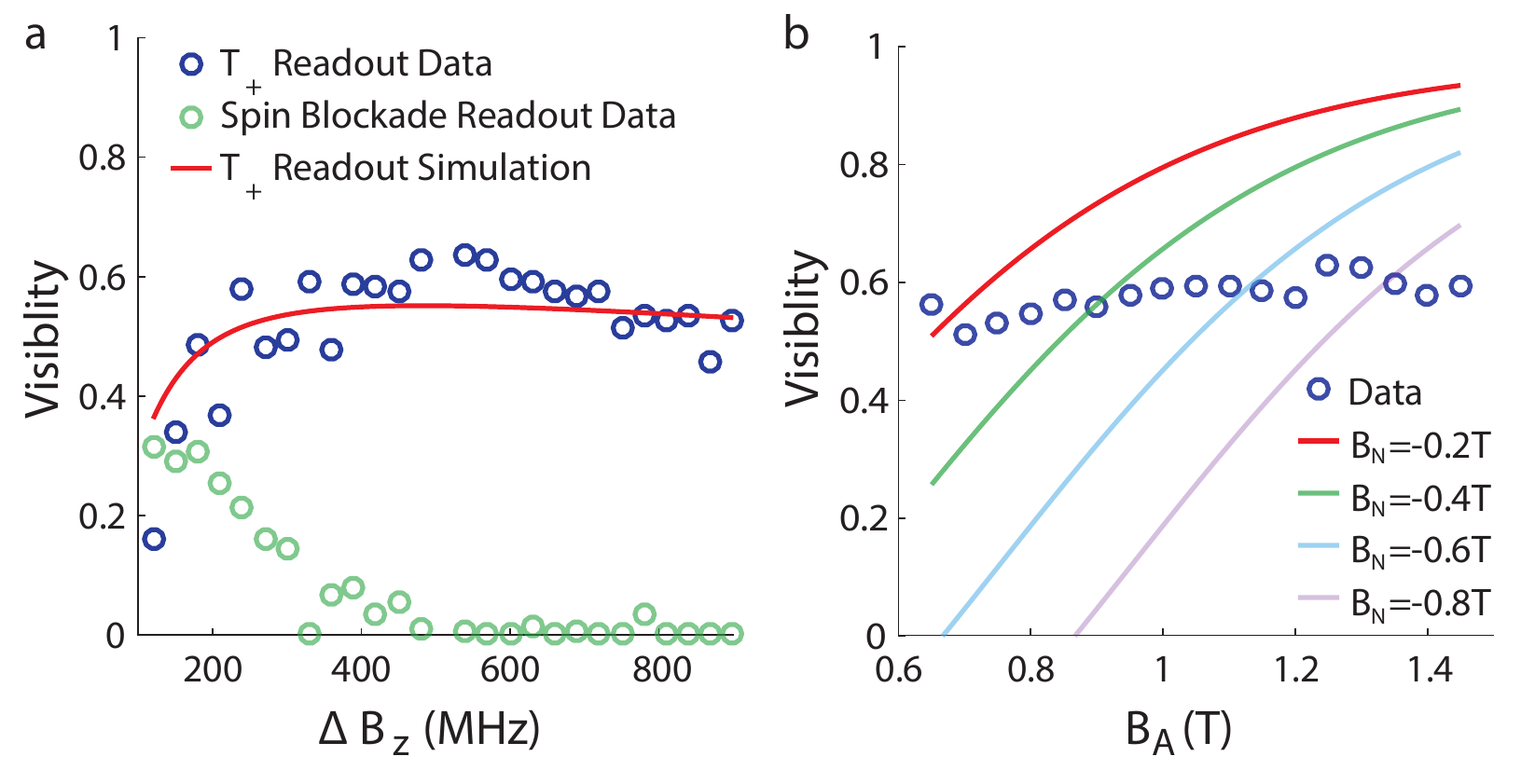}%
	\caption{(a) Measurements of the visibility of the spin blockade and T$_+$ readout methods as a function of $\Delta B_{z}$.  Red curve is a simulation of the T$_+$ method.  The visibility of the T$_+$ method is superior at large $\Delta B_z$.  (b) Measurement and simulations with varied $B_N$ of the visibility of the T$_+$ method as a function of $B_{A}$.  The data does not follow one simulation curve, suggesting that the $B_N$ produced by DNPS is a function of $B_A$.}
\end{figure}

We determined the contrast of the readout methods that we tested by finding the measurement fidelity\cite{Barthel2} for the ground, F$_G$ and excited states, F$_E$, as detailed in the Supplementary Materials.  We used these quantities to calculate the visibility, given by F$_{G}$+F$_{E}$-1.  In Figure 3a we present the measured visibility of spin blockade and T$_{+}$ readout techniques as a function of $\Delta B_{z}$.  In Figure 3a we also present a simulation for the visibility of the T$_{+}$ readout and note the agreement with the data.  

We see that the spin blockade readout visibility decreases very quickly with increasing $\Delta B_z$ as we expect from the increasing decay rate from T$_{0}$(1,1) to S(0,2) at the measurement point.  The T$_{+}$ readout is poor at small $\Delta B_{z}$ because J$(\epsilon)$ is comparable to $\Delta B_{z}$ which gives both qubit states the ability to decay to $\uparrow$S(1,2).  However, the T$_+$ method has large visibility for $\Delta B_z>$200 MHz.  We note also the slow fall off of visibility for $\Delta B_{z}>$500 MHz.  This is due to $\Delta B_{z}$ decreasing the energy splitting between the $\uparrow\downarrow$ state and the $\uparrow\uparrow$ state, decreasing the thermodynamic equilibrium occupation of $\uparrow\uparrow$, as can be seen from the energies given in the Supplementary Materials.  Flipping the direction of $\Delta B_{z}$ would give a weak improvement instead because $\Delta B_{z}$ would increase the energy difference between $\uparrow\uparrow$ and $\uparrow\downarrow$ rather than decrease it.  We compare the performance of the T$_{+}$ and another previously published readout method\cite{Studenikin} as a function of $\Delta B_{z}$ in the Supplementary Materials.

In Figure 3b we present the data for the T$_{+}$ readout method visibility versus the applied magnetic field, $B_{A}$.  We find only a weak dependence on the $B_{A}$ while the model predicts a sharp increase.  Past measurements have shown that DNP pumps both the difference field, $\Delta B_z$, and sum field, $B_N$ experienced quantum dots due to the polarized nuclei.  The magnetic field experienced by the qubit is \begin{math}B=B_{A}+B_N \end{math}.  Pumping with T$_+$ states flips nuclei such that $B_N<$0 while pumping with S states yields $B_N>$0.  While measuring the data presented in Figure 3b, we observed increasing DNP times required for a given value of $\Delta B_z$ to the extent that it took 10 times longer to stabilize $\Delta B_{Z}$ at $B_{A}$=1.4 T than at $B_{A}$=0.7 T.  This suggests that nuclei are flipped more symmetrically between the dots with increasing $B_A$, yielding larger magnitude $B_N$, because DNP is less efficient at pumping $\Delta B_z$.  In Figure 3b we plot simulations at several different $B_N$ and see that the data transition between curves with increasingly negative $B_N$, consistent with DNP becoming less effective at generating $\Delta B_{z}$ at larger $B_{A}$.  The magnetic field dependence of DNP pumping rates of $\Delta B_z$ and $B_N$ is a subject of current investigation.

\begin{figure}
	\includegraphics[scale=.52]{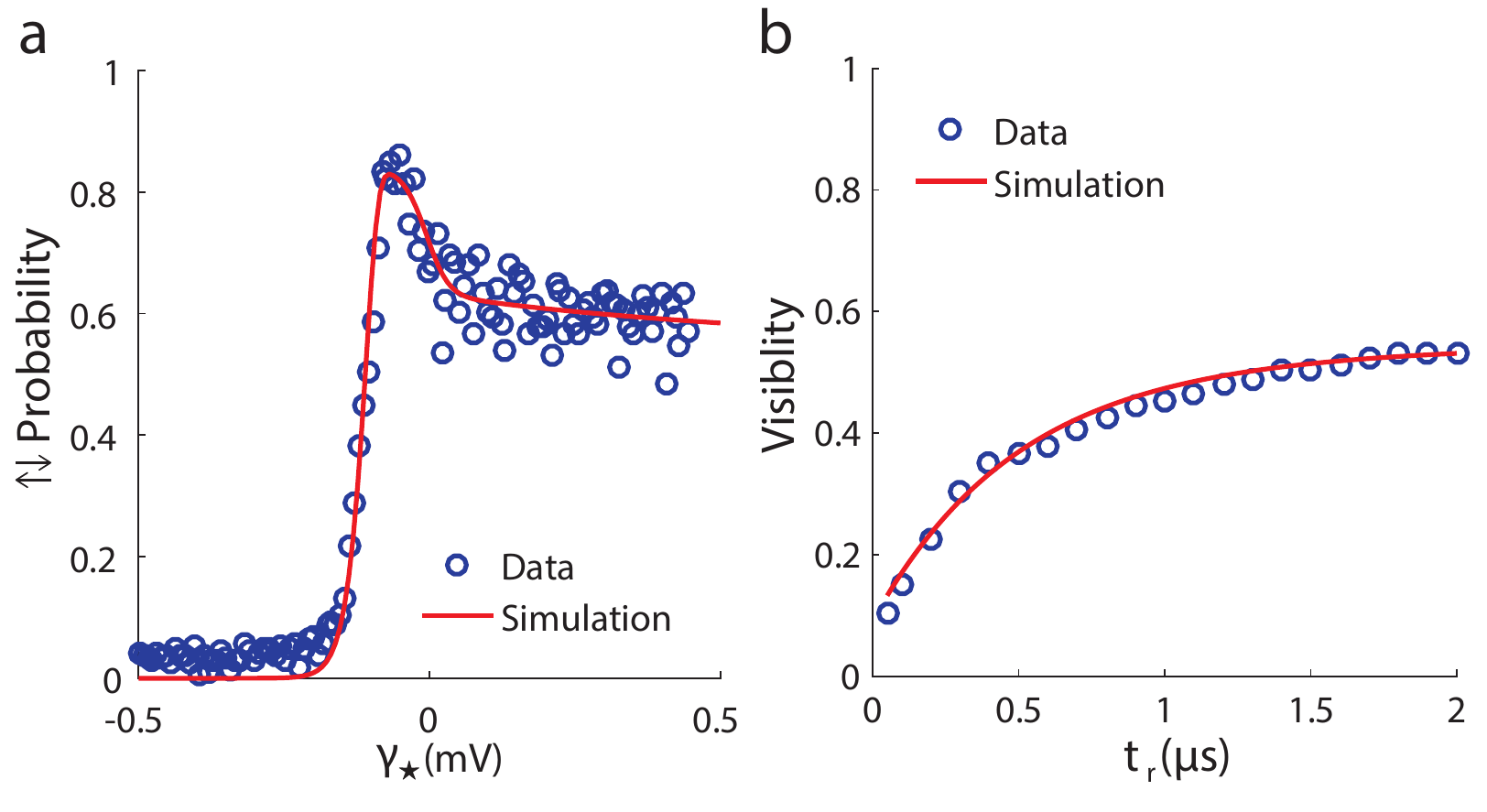}%
	\caption{(a) Measurements and simulation for the probability that the $\uparrow\downarrow$ state correctly transitions as a function of $\gamma_{\star}$.  The peak occurs where the required transitions are energetically favorable while still keeping undesirable transitions unfavorable.  (b) Measurement and simulation of the visibility of the T$_+$ as a function of t$_{r}$.  Longer times allow the qubit to more completely transition to the desired end state.}
\end{figure}

The fidelity of the T$_+$ readout method depends strongly on the readout position because the technique relies on the desired transitions being energetically favorable while the undesired transitions remain unfavorable.  The energy spectrum of available states as a function of $\gamma$ is shown in Figure 1e.  We select the optimal readout position by repeatedly preparing $\uparrow\downarrow$(1,1) and immediately attempting to measure at readout positions with different $\gamma_{\star}$, as shown in Figure 2d.  We plot data and a simulation of the probability that the measurement correctly identified the $\uparrow\downarrow$ state in Figure 4a.

When $\gamma_{\star}\ll$0 the $\uparrow$S(1,2) state has far more energy than $\uparrow\downarrow$(1,1), preventing the first transition required for T$_{+}$ readout.  As $\gamma_{\star}$ approaches zero $\uparrow$S(1,2) comes into resonance with $\uparrow\downarrow$(1,1) and we see a dramatic upturn in the probability of transitioning because there are thermally excited electrons that allow for the first transition.  When $\gamma_{\star} >$0 the probability drops again because the desired end state, $\uparrow\uparrow$(1,1) is not the lowest in energy during the readout process so it is not the most thermodynamically populated.  All other measurements in this paper were performed at the optimal measured readout position.

To optimize the T$_{+}$ readout, we also investigated the dependence of the visibility on t$_r$.  Our simulations and experiments showed little dependence on how quickly $\gamma$ was increased to ramp the voltages from Point A to Point B, where $\gamma=\gamma_{\star}$, but a strong dependence on the time, t$_{r}$, over which $\gamma$ was varied to change the voltages back from Point B back to Point A.  We present measurements and simulations for the visibility as a function of t$_{r}$ in Figure 4b.  The visibility sharply improves with increasing t$_{r}$ because the qubit has time to equilibrate as $\gamma$ is varied resulting in a higher occupation of T$_+$.  At very short times, ($\uparrow$,S) is rapidly raised above $\uparrow\downarrow$ state, allowing for undesirable transitions and reducing visibility.

The maximum visibility that we observe is approximately 0.6, corresponding to an average readout fidelity of 80$\%$.  This is limited by the equilibrium thermodynamic occupation of the states that the qubit transitions through during the shelving process.  This thermodynamic limit can be improved by decreasing the electron temperature or by using $\Delta B_z$ and $B$ to increase the energy splittings between the states used for shelving.  As mentioned above, the direction of $\Delta B_z$ can be chosen so that it increases the relevant splittings.  While the direction of $\Delta B_z$ in these experiments was governed by using DNP and decreased the relevant splittings, the direction is more flexible when generated by a micromagnet\cite{Pioro, Takeda, Wu, Kawakami} so that visibility can instead be enhanced. Another benefit of using a micromagnet is that $B_N$ will remain fixed, so that we have direct control of $B$ through $B_A$.  We expect to observe the behavior predicted by the simulations in Figure 3b, allowing this method to achieve visibilities above 90$\%$ by increasing $B_A$.  

%\section{Conclusion}
We demonstrated that the T$_{+}$ readout method allows for measurements with large $\Delta B_{z}$, a regime that was previously inaccessible due to low readout visibility.  We have also demonstrated that calibrating t$_{r}$ and $\gamma_{\star}$ is critical to optimizing the visibility.  Additionally, we have identified that using an external source of $\Delta B_z$, such as a micromagnet, should enable higher fidelity readout by the application of larger $B_A$ and prudently selecting the direction of $\Delta B_z$.  We expect that these changes should enable visibilities in line with other high quality qubit readouts.  The T+ readout technique is also applicable to scalable architectures that map a single spin qubit onto S-T states for readout\cite{Fogarty}.  

The concept of using a shelving step before measurement is relevant to any system where readout is limited by decay processes during measurement.  We have demonstrated that visibilities can be increased by transferring the qubit into states that are immune to the decay pathway before measurement.  We have also developed a method for simulating processes that rely on spin dependent tunneling between a quantum dot and a reservoir. This can be used to optimize the initialization and readout in a wide variety of qubits because they rely on these tunneling processes.  Our demonstration of using experiments and simulations to develop the T$_+$ readout method can serve as a guide for other researchers who need to develop readout schemes tailored to their specific experimental requirements.

%\section{Acknowledgments}
Work at Harvard was funded by Army Research Office grants W911NF-15-1-0203 and W911NF-17-1-024. Samples were fabricated at the Harvard University Center for Nanoscale Systems (CNS), a member of the National Nanotechnology Infrastructure Network (NNIN), which is supported by the National Science Foundation under NSF award No. ECS0335765.  Work at Purdue was supported by the Department of Energy, Office of Basic Energy Sciences, under Award number DE-SC0006671. Additional support for the MBE growth facility from the W. M.  Keck Foundation and Nokia Bell Laboratories is gratefully acknowledged.  S.P.H. was supported by the Department of Defense through the National Defense Science Engineering Graduate Fellowship Program. L.A.O. was supported by the Army Research Office through the Quantum Computing Graduate Research Fellowship Program.

\setcounter{figure}{0}
\renewcommand{\thefigure}{S\arabic{figure}}
\newpage

\section{Supplementary Materials}

\subsection{Experimental}

Our devices were fabricated on GaAs/AlGaAs heterostructure with the 2DEG located 91 nm below the surface.  At 4K the electron density n=1.5 $\times$ $10^{11}$ cm$^{-2}$ and mobility $\mu$=2.5 $\times$ $10^{6}$ cm$^{2}$V$^{-1}$s$^{-1}$.  Measurements presented in this work were taken in a dilution refrigerator with a base temperature of approximately 15 mK.

The direction of $\Delta B_{z}$ in these experiments was dictated by DNP's ability to more strongly pump with T$_+$ than with S.  Pumping with T$_+$ flips spins counter to the applied magnetic field, reducing the Zeeman splitting experienced by the qubits.  This brings the avoided crossing between T$_+$ and the ground state S deeper into the (1,1) region.  The hyperfine term used for pumping only couples T$_+$(1,1) to S(1,1), meaning that pumping is more effective when the ground state singlet branch has more weight in S(1,1) than S(0,2).  Pumping with S achieves the opposite effect, increasing the Zeeman splitting and decreasing the coupling between the ground state singlet branch and T$_+$ and reducing the ability to pump the gradient.

In order to measure the readout fidelity we need to prepare and measure both the excited and ground states.  We prepare the ground state by loading two electrons into the double dot deep in (0,2), where S(0,2) is the ground state, and then adiabatically ramp into (1,1) where $\uparrow\downarrow$ is the ground state.  We prepare the excited state by first preparing the ground state in (1,1) and then rotating it by using a sinusoidally time varying $\epsilon$ to generate a Rabi $\pi$ pulse about J.  We have previously reported 99$\%$ fidelity randomized benchmarking\cite{Emerson, Dankert, Levi} measurements of $\pi$ pulses of  at $\Delta B_{z}$=900 MHz\cite{Nichol}.  While the $\pi$ pulse will become slightly worse at lower gradients, these state preparation errors are small compared to our readout errors and only negligibly affect our results.  Figure S1a shows histograms of measurements of the excited and ground state of the qubit that were used to determine the readout visibility at $\Delta B_z$ = 900 MHz.

\begin{figure}
	\includegraphics[scale=.5]{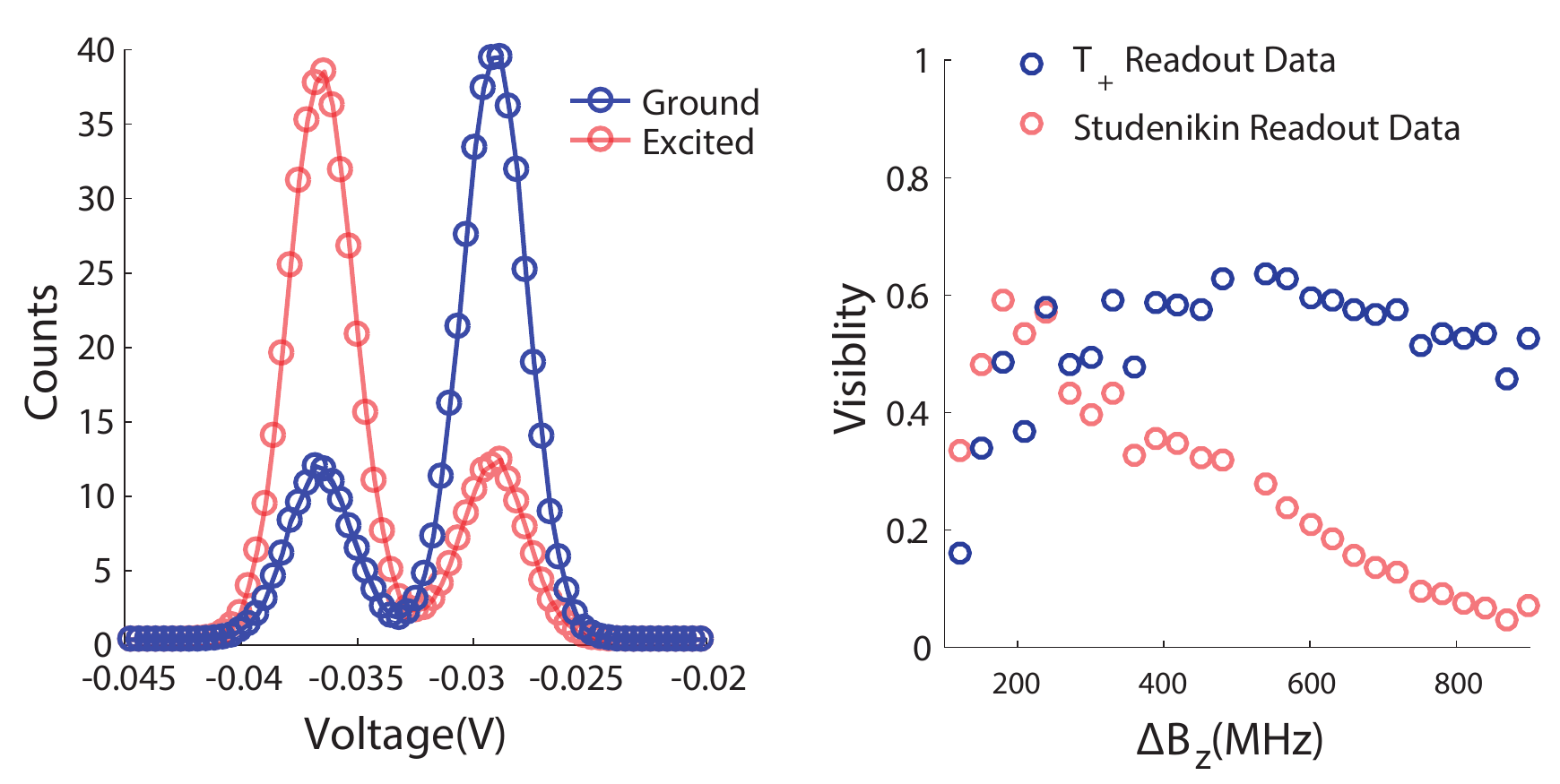}%
	\caption{(a) Histograms of measurements using the T$_+$ method to readout the qubit's ground and excited state.  Histograms are used to determine the readout visibility, in this case for $\Delta B_z$=900 MHz. (b) A comparison of the performance of the T$_+$ and Studenikin readout methods versus $\Delta B_z$.  Studenikin is preferable for $\Delta B_{z}<$300 MHz while the T$_+$ has greater visibility for $\Delta B_{z}>$400 MHz. }
\end{figure}

For thoroughness, we also discuss the readout method described by Studenikin\cite{Studenikin} that improves readout contrast by choosing the measurement point in (0,2) so that there is a (1,2) charge state with lower energy than the T$_{0}$(1,1) state but still higher in energy than the S(0,2) state.  The right dot of the double quantum dot is coupled to the lead, allowing a third electron to tunnel into double dot when the qubit is in T$_{0}$(1,1) state because it is an energetically favorable transition but not for when the qubit is in S(0,2).  This improves contrast because the charge sensor now detects different numbers of electrons for the two qubit states as opposed to just different positions of two electrons for both states like the spin blockade readout method.  This method also suffers as $\Delta B_{z}$ increase because the transitions from T$_{0}$(1,1) to the excited S(1,1) compete with the transitions of T$_{0}$(1,1) to (1,2) and decrease measurement contrast as $\Delta B_{z}$ increases, as in the spin blockade readout technique. We note that the Studenikin readout method works up to larger gradients than spin blockade because there is a competition between the decays of T$_{0}$ to the desired $\uparrow$S(1,2) and S(0,2) but that the undesirable transitions dominates as $\Delta B_{z}$ surpasses 400 MHz, as shown in Figure S1b.  

We also acknowledge two other recently published S-T readout mechanisms that, like the Studenikin method, rely on transitions through a (1,2) state in the region where (0,2) is the ground state.  The Broome method\cite{Broome} addresses how to readout a S-T qubit in which both quantum dots are equally coupled to the charge sensor while the Fogarty method\cite{Fogarty} maps a single electron spin in a large array onto the S-T basis for readout.  While these methods solve the problems that they were intended to address, we expect that they should also suffer in large magnetic field gradients for the same reason as the Studenikin method.

\subsection{Theoretical}
The T$_{+}$ readout technique relies on electrons entering and leaving the quantum dot so that the qubit can relax to lower energy states.  To model this, we treat the tunneling term, $\tau$, as a small perturbation to the Hamiltonian that confines the electrons in the quantum dots so that $\tau$ couples the (1,2) states to the (1,1) states.  Because we only allow for tunneling into the right quantum dot, $\tau$ can only mix states which have the same spin in the left dot, as shown below:

Basis=$\begin{bmatrix}
	\downarrow\downarrow \\
	\downarrow S \\
	\downarrow\uparrow \\
	\uparrow\downarrow \\
	\uparrow S \\ 
	\uparrow\uparrow \\
\end{bmatrix}$, 
$\tau=\tau_{0}
\begin{bmatrix}
	0 & 1 & 0 & 0 & 0 & 0\\
	1 & 0 & 1 & 0 & 0 & 0\\
	0 & 1 & 0 & 0 & 0 & 0\\
	0 & 0 & 0 & 0 & 1 & 0\\
	0 & 0 & 0 & 1 & 0 & 1\\
	0 & 0 & 0 & 0 & 1 & 0
\end{bmatrix}$

The term $\tau_{0}$ controls the strength of the tunneling interaction and should be the same constant between all (1,1) and (1,2) states\cite{Bardeen}.  

In the real experiment, the qubit eigenstates are not perfectly $\uparrow\downarrow$ and $\downarrow\uparrow$ because $J \neq 0$.  This means that the ground and excited states take the following forms, where $\phi$ is defined as tan$(\phi)=J/\Delta B_{z}$.

\begin{align*}
\ket{G}=\cos(\frac{\phi}{2})\ket{\uparrow\downarrow}-\sin(\frac{\phi}{2})\ket{\downarrow\uparrow}\\
\ket{E}=\sin(\frac{\phi}{2})\ket{\uparrow\downarrow}+\cos(\frac{\phi}{2})\ket{\downarrow\uparrow}
\end{align*}

To simplify notation, we number the relevant states in the following way:  $\uparrow\uparrow$=1, $\uparrow$S=2, G=3, E=4, $\downarrow$S=5 and $\downarrow\downarrow$=6.  The states and the transitions, $\Gamma_{ij}$, are shown in Figure S2.  The states have the following energies, $E_{i}$

\begin{align*}
E_{1} &=-g\mu_{B}B_{0}\\
E_{2} &=-g\mu_{B}B_{0}+K(\gamma)\\
E_{3} &=-\Omega/2\\
E_{4} &=\Omega/2\\
E_{5} &=\frac{1}{2}g\mu_{B}\Delta B_{z}+K(\gamma)\\
E_{6} &=g\mu_{B}B_{0}
\end{align*}
The (1,2) states, 2 and 5, have an additional energy the $K(\gamma)$, the energy difference between the $\uparrow\uparrow$ and $\uparrow$S states, because $\gamma$ controls the energy of the (1,2) states relative to the (1,1) states.  It is defined so that $K(\gamma)\propto -\gamma$ because K is positive in (1,1) where $\gamma$ is negative.  Some care must be given when considering the energy difference between quantum dot states with numbers of electrons.  Our choice of K($\gamma$)=0 at the charge transition allows us to set the chemical potential equal to zero because at this point an electron can tunnel from the fermi level into the quantum dot without paying an energy cost.

\begin{figure}
	\includegraphics[scale=.6]{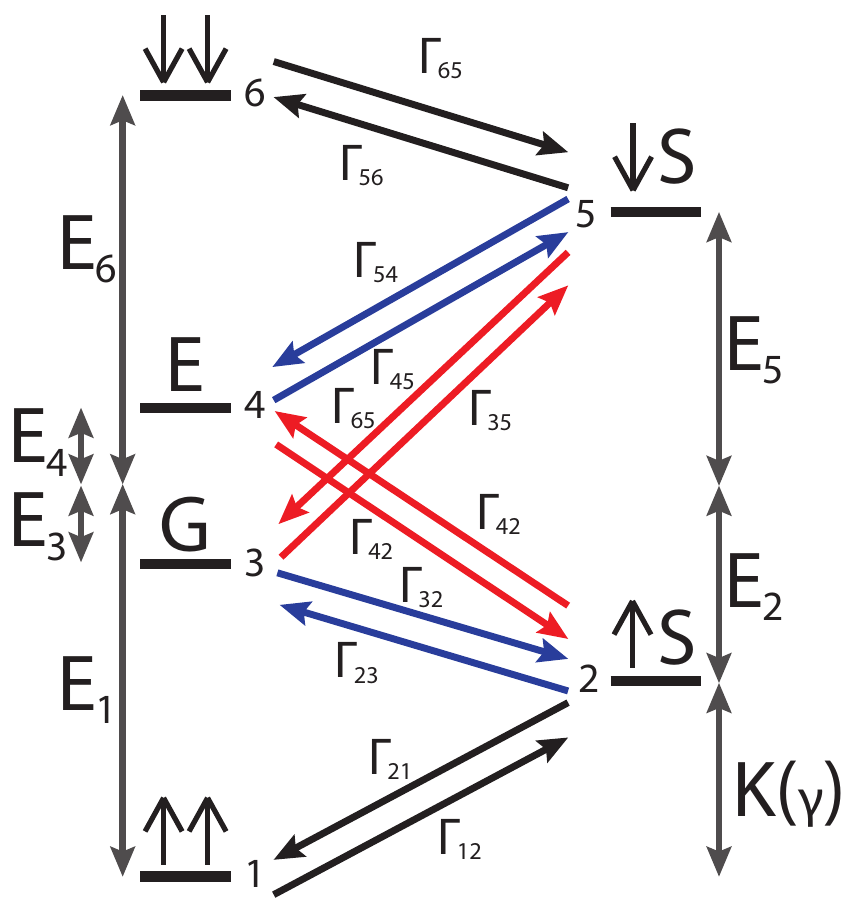}%
	\caption{Diagram of the energies of the states used for readout.  The single headed arrows show the allowed transitions between states.  Those in black have not dependence on $\phi$, those in blue scale as $\cos^{2}(\phi/2)$ and those in red scale as $\sin^{2}(\phi/2)$.}
\end{figure}

From Fermi's Golden Rule we have equations (1) and (2) in the text.  We can simplify these by using $\Gamma_{0}=\frac{2\pi}{\hbar}\tau\rho_{f}$.

\begin{align*}
\Gamma_{ij} &=\Gamma_{0}\alpha_{ij}(\phi)f(\Delta E_{ij},T,\mu)\\
\Gamma_{ji} &=\Gamma_{0}\alpha_{ij}(\phi)(1-f(-\Delta E_{ji},T,\mu))
\end{align*}

\begin{figure*}
	$R=\begin{bmatrix}
	-\Gamma_{12} & \Gamma_{21} & 0 & 0 & 0 & 0\\
	\Gamma_{12} & -\Gamma_{21}-\Gamma_{23}-\Gamma_{24} & \Gamma_{32} & \Gamma_{42} & 0 & 0\\
	0 & \Gamma_{23} & -\Gamma_{32}-\Gamma_{35} & 0 & \Gamma_{53} & 0\\
	0 & \Gamma_{24} & 0 & -\Gamma_{42}-\Gamma_{45} & \Gamma_{54} & 0\\
	0 & 0 & \Gamma_{35} & \Gamma_{45} & -\Gamma_{53}-\Gamma_{54}-\Gamma_{56} & \Gamma_{65}\\
	0 & 0 & 0 & 0 & \Gamma_{56} & -\Gamma_{65}
	\end{bmatrix}$
\end{figure*} 

The terms $\alpha_{ij}=\alpha_{ji}$ are the overlap of the spin states and the nonzero terms are given below.

\begin{align*}
\alpha_{12}&=1\\	
\alpha_{32}&=\cos^{2}(\phi/2)\\	
\alpha_{42}&=\sin^{2}(\phi/2)\\
\alpha_{35}&=\sin^{2}(\phi/2)\\	
\alpha_{45}&=\cos^{2}(\phi/2)\\	
\alpha_{65}&=1
\end{align*}

Knowing the transition rates allows us to calculate the probability, P$_{i}$ (P$_{j}$), that the qubit is in state $i$ ($j$) by solving the following six coupled linear first order differential equations for all six states shown in Figure S2.

\begin{align*}
\frac{dP_{i}}{dt}=\sum_{j=2,5}(-P_{i}\Gamma_{ij}+P_{j}\Gamma_{ji})\\ \frac{dP_{j}}{dt}=\sum_{i=1,3,4,6}(-P_{j}\Gamma_{ji}+P_{i}\Gamma_{ij})
\end{align*}

The first term on the right hand side states that the probability that the qubit remains in state i decreases because it can transition to state j at a rate that is proportional to how likely the qubit is in state i to begin with and the tunneling rate $\Gamma_{ij}$.  The second term is the opposite, stating that the rate of transitioning to i increases when the other states, j, are more occupied.  

We solve these coupled differential equations by turning them into a matrix equation which then has the form $\frac{d}{dt}\vec{p}$=R$\vec{p}$ ⃑where $\vec{p}$ is the vector of $P_{i}$'s and $P_{i}$'s and R is a matrix containing all the $\Gamma_{ij}$, shown at the top of the page.  We project $\vec{p}$ onto the eigenvectors of R, $\vec{v}_{k}$, with eigenvalues $a_{k}$, whose time dependence is given by $\vec{v}_{k}(t)=e^{a_{k}t}\vec{v}_{k}$.  The time evolution of $\vec{p}$ is then just the sum of the time evolutions of the projections onto the eigenvectors.  We note that one eigenvector will always be the thermal equilibrium because our rates are inherently thermodynamic due to their dependence on temperature through the Fermi-Dirac distribution.  All other solutions will decay to zero with time scales that depend on the $\Gamma_{ij}$'s.

In our experiment, $\gamma$, is a function of time because the qubit is biased from deep in (1,1) to the transition to (1,2).  We simulate this by discretizing time and assuming that for each time step of length $\Delta$t, $K(\gamma)$ has a constant value $K(\gamma(t))$ and that this constant value controls the rates of decay.  We make sure that $\Delta$t$\ll\Gamma_{0}^{-1}$ so that time evolution at each step is small.  We initialize $\vec{p}$(0) with all the weight in either the ground or excited qubit state.  At each time step we project $\vec{p}$(t) onto the eigenvectors of R(t) and evolve it for time $\Delta$t which yields $\vec{p}$(t+$\Delta$t).  We repeat this for each time step to find the total time evolution of our qubit.  We have performed these computations while varying the same parameters as have experimentally investigated by changing the state energies and $\gamma$(t).

All simulations were performed with T=90 mK, J=50 MHz, $\Delta$t=1 ns, $\Gamma_{0}$= 7.143$\times$10$^7$ s$^{-1}$ and $\mu$=0 meV.  Unless otherwise specified $\Delta B_{z}$=900 MHz, $B_{A}$=0.7 T, t$_{r}$=2 $\mu$s and $\gamma_{\star}$=-.065 mV.  For the simulations with $\Delta B_{z}$ and t$_{r}$ varied, $B_N$=-.23 T.  For the simulation with $\gamma_{\star}$ varied $B_N$= -.12 T which is not unreasonable given that this data set was taken several months before the others when the tuning parameters were different.  The individual curves are labeled in the figure for $B_{A}$ varied.

\end{document}